\documentclass[conference]{IEEEtran}
\IEEEoverridecommandlockouts

\usepackage{cite}
\usepackage{amsmath,amssymb,amsfonts}
\usepackage{algorithmic}
\usepackage{graphicx}
\usepackage{textcomp}
\usepackage{xcolor}
\usepackage{tikz}
\usepackage{array}
\usepackage{multirow}
\usepackage{booktabs}
\usepackage[utf8]{inputenc}
\usepackage{textcomp}
\usetikzlibrary{shapes.geometric, arrows, positioning, fit, backgrounds}

\def\BibTeX{{\rm B\kern-.05em{\sc i\kern-.025em b}\kern-.08em
    T\kern-.1667em\lower.7ex\hbox{E}\kern-.125emX}}

\usepackage[utf8]{inputenc}
\begin{document}

\title{Evolution of Buffer Management in Database Systems: From Classical Algorithms to Machine Learning and Disaggregated Memory}

\author{
\IEEEauthorblockN{Prudhvi Gadupudi}
\IEEEauthorblockA{\textit{Department of Computer Science and Engineering} \\
\textit{The Pennsylvania State University}\\
University Park, PA 16802, USA \\
gvp5349@psu.edu}
\and
\IEEEauthorblockN{Suman Saha}
\IEEEauthorblockA{\textit{Department of Computer Science and Engineering} \\
\textit{The Pennsylvania State University}\\
University Park, PA 16802, USA \\
sumsaha@psu.edu}
}

\maketitle

\begin{abstract}
Buffer management remains a critical component of database and operating system performance, serving as the primary mechanism for bridging the persistent latency gap between CPU processing speeds and storage access times. This paper provides a comprehensive survey of buffer management evolution spanning four decades of research. We systematically analyze the progression from foundational algorithms like LRU-K, 2Q, LIRS, and ARC to contemporary machine learning-augmented policies and disaggregated memory architectures. Our survey examines the historical OS-DBMS architectural divergence, production system implementations in PostgreSQL, Oracle, and Linux, and emerging trends including eBPF-based kernel extensibility, NVM-aware tiering strategies, and RDMA-enabled memory disaggregation. Through analysis of over 50 seminal papers from leading conferences (SIGMOD, VLDB, OSDI, FAST), we identify key architectural patterns, performance trade-offs, and open research challenges. We conclude by outlining a research direction that integrates machine learning with kernel extensibility mechanisms to enable adaptive, cross-layer buffer management for heterogeneous memory hierarchies in modern database systems.
\end{abstract}


\section{Introduction}

The memory hierarchy has been a fundamental aspect of computer system design for over half a century. The persistent disparity between CPU processing speeds and storage access latencies necessitates intelligent buffer management: maintaining a carefully selected subset of data in fast, volatile memory to service repeated accesses efficiently. In both Database Management Systems (DBMS) and Operating Systems (OS), buffer management transcends simple caching—it is the primary determinant of overall system throughput and response latency.

The landscape of buffer management has undergone significant transformation. Early systems relied on simple heuristics like LRU (Least Recently Used), which assumed uniform access costs and homogeneous storage media \cite{stonebraker1981}. Contemporary systems must navigate a complex multi-tiered memory hierarchy encompassing processor caches, DRAM, CXL-attached memory, persistent memory (NVM), and various storage technologies (NVMe SSD, HDD), each with distinct latency, bandwidth, and cost characteristics \cite{zhou2021spitfire, zhong2024memstrata}.

This paper presents a comprehensive survey of buffer management research, organized chronologically and thematically. Our survey makes the following contributions:

\begin{enumerate}
\item \textbf{Historical Analysis:} We trace the evolution from Stonebraker's 1981 critique of OS support for databases through the development of sophisticated replacement algorithms (LRU-K, LIRS, ARC) to modern learned policies.

\item \textbf{Systematic Taxonomy:} We classify buffer management approaches across multiple dimensions: algorithmic complexity, scan resistance, self-tuning capability, hardware awareness, and deployment context (Tables I-III).

\item \textbf{Implementation Study:} We examine how production systems (PostgreSQL, Oracle, MySQL, Linux) implement buffer management, analyzing their concurrency mechanisms, scalability characteristics, and practical trade-offs.

\item \textbf{Emerging Trends:} We survey recent advances in machine learning-based policies, eBPF kernel extensibility, NVM-aware algorithms, and disaggregated memory architectures from 2020-2025 literature.

\item \textbf{Research Directions:} We identify open challenges and outline promising directions for future work, including cross-layer optimization through eBPF and ML-based adaptation.
\end{enumerate}

The remainder of this paper is organized as follows: Section II examines foundational work establishing the OS-DBMS architectural divide and economic models. Section III surveys classical replacement algorithms. Section IV analyzes production system implementations. Section V reviews cost-aware policies for modern storage media. Section VI examines machine learning approaches. Section VII explores eBPF-based kernel extensibility. Section VIII discusses disaggregated memory systems. Section IX presents research directions, and Section X concludes.

\section{Historical Foundation and Economic Models}

\subsection{The OS-DBMS Architectural Divergence}

The architecture of modern database buffer management was profoundly influenced by Stonebraker's 1981 paper "Operating System Support for Database Management" \cite{stonebraker1981}. Stonebraker argued that general-purpose operating systems make fundamentally suboptimal decisions for database workloads. The OS manages memory with "blind" global policies, typically LRU variants, designed for fairness across diverse processes without application-specific semantic knowledge.

In contrast, a DBMS possesses rich semantic information about its access patterns. Consider a sequential scan of a multi-gigabyte table: the DBMS understands that each page will be accessed exactly once and never revisited, yet a naive OS policy would cache these pages, potentially evicting highly valuable B-tree index nodes in a phenomenon known as cache pollution. This fundamental mismatch led to the "double buffering" problem—data residing redundantly in both the OS page cache and the DBMS user-space buffer pool, wasting memory and introducing synchronization overhead.

This insight drove modern database systems (PostgreSQL, Oracle, MySQL) to bypass OS caching mechanisms using flags like O\_DIRECT on Linux or FILE\_FLAG\_NO\_BUFFERING on Windows. By managing their own memory pools, databases can implement sophisticated, workload-aware eviction policies, but at the cost of reimplementing memory management logic and losing potential OS-level optimizations \cite{interdb2024}.

\subsection{Economic Foundations: The Five-Minute Rule}

While Stonebraker established the architectural principles, Gray and Putzolu provided the economic framework. Their 1987 "Five-Minute Rule" offered a quantitative method for buffer sizing decisions \cite{gray1987}. The rule states that for a page of size $P$ accessed with frequency $f$, it should be kept in memory if:

\begin{equation}
f > \frac{C_{memory}}{C_{disk\_access}}
\end{equation}

where $C_{memory}$ is the cost of storing one page in RAM and $C_{disk\_access}$ is the cost of one disk I/O operation. For the technology of 1987, this threshold occurred at approximately one access every five minutes for a 1KB page.

The rule has proven remarkably durable. Gray and Graefe revisited it in 1997, finding the five-minute threshold still held despite a decade of hardware evolution \cite{gray1997}. Graefe updated the rule in 2008 for Flash memory, establishing a two-tier hierarchy: pages accessed every 5 minutes should stay in DRAM, while those accessed every 2 hours could reside on Flash rather than HDD \cite{graefe2008}.

In today's landscape with tiered memory (DRAM at \$5/GB, CXL at \$2/GB, NVMe at \$0.10/GB), the Five-Minute Rule's underlying principle remains relevant, though the economics have shifted toward bandwidth costs and latency-sensitive workloads rather than mechanical seek times \cite{zhong2024memstrata}.

\begin{table*}[t]
\caption{Comparative Analysis of Classical Buffer Replacement Algorithms}
\label{tab:classical}
\centering
\begin{tabular}{p{1.8cm}p{2.2cm}p{1.5cm}p{2cm}p{2cm}p{3cm}p{2cm}}
\toprule
\textbf{Algorithm} & \textbf{Core Metric} & \textbf{Complexity} & \textbf{Scan Resistant} & \textbf{Self-Tuning} & \textbf{Key Innovation} & \textbf{Primary Use} \\
\midrule
LRU & Recency & O(1) & No & No & Simple stack discipline & Legacy systems \\
\midrule
LFU & Frequency & O(log N) & Yes & No & Access count tracking & Static workloads \\
\midrule
LRU-K \cite{oneil1993lruk} & K-distance & O(log N) & Yes & No & Correlated reference period & Commercial DBMS \\
\midrule
2Q \cite{johnson19942q} & Queue membership & O(1) & Yes & No & Probationary queue & PostgreSQL (influenced) \\
\midrule
LIRS \cite{jiang2002lirs} & Inter-reference recency & O(1) & Yes & No & IRR vs. recency distinction & MySQL 8.0 \\
\midrule
ARC \cite{megiddo2003arc} & Adaptive recency/frequency & O(1) & Yes & Yes & Ghost cache adaptation & ZFS, storage arrays \\
\midrule
CLOCK-Pro \cite{jiang2005clock} & LIRS approximation & O(1) & Yes & No & Hardware-friendly CLOCK & Linux kernel (influenced) \\
\midrule
CAR \cite{bansal2004car} & ARC variant & O(1) & Yes & Yes & Patent-free ARC alternative & Open-source systems \\
\bottomrule
\end{tabular}
\end{table*}

\section{Classical Replacement Algorithms}

This section surveys the "golden age" of buffer replacement algorithms (1990-2005), which produced the foundational policies still powering most production systems today.

\subsection{LRU-K: Temporal Pattern Recognition}

O'Neil et al.'s 1993 LRU-K algorithm marked a significant advancement beyond simple LRU \cite{oneil1993lruk}. Rather than using only the most recent access timestamp ($K=1$), LRU-K tracks the last $K$ access times for each page. The eviction criterion is the "backward K-distance"—the time elapsed since the $K^{th}$ most recent access.

The key insight is the concept of "Correlated Reference Period": multiple accesses occurring within a short temporal window should be treated as a single utilization event. For $K=2$ (LRU-2), pages accessed only once (e.g., during sequential scans) have infinite backward-2 distance, making them immediate eviction candidates. This provides robust scan resistance while preserving pages with sustained access frequency.

The algorithm distinguishes between transient locality (brief bursts) and persistent locality (sustained access patterns). However, LRU-K's $O(\log N)$ complexity for maintaining the priority queue makes it computationally expensive for high-throughput systems \cite{oneil1993lruk}.

\subsection{2Q: Efficient Approximation}

Johnson and Shasha addressed LRU-K's computational overhead with 2Q (Two Queue), achieving $O(1)$ complexity while approximating LRU-2 behavior \cite{johnson19942q}. The buffer is partitioned into three logical sections:

\begin{itemize}
\item $A_m$ (Main Queue): An LRU list containing hot pages
\item $A_{1in}$ (First-In): Newly accessed pages (resident in memory)
\item $A_{1out}$ (First-Out): Ghost entries tracking recently evicted pages
\end{itemize}

New pages enter $A_{1in}$ on first access. If evicted from $A_{1in}$ without re-reference, the page is removed from memory but its identifier is retained in $A_{1out}$. A subsequent access to a page in $A_{1out}$ triggers promotion to $A_m$. This probationary mechanism efficiently filters one-time accesses (sequential scans) while identifying and promoting frequently accessed pages. The simplicity and effectiveness of 2Q influenced PostgreSQL's buffer management design \cite{johnson19942q}.

\subsection{LIRS: Redefining the Metric}

Jiang and Zhang's 2002 LIRS (Low Inter-reference Recency Set) algorithm represented a paradigm shift \cite{jiang2002lirs}. Rather than using absolute recency (time since last access), LIRS employs Inter-Reference Recency (IRR)—the temporal distance between the last two accesses to a page.

A page with small IRR exhibits strong temporal locality and should be retained. A page with large IRR (infrequent re-reference) becomes an eviction candidate, even if its absolute recency is recent. LIRS maintains two data structures:

\begin{itemize}
\item Stack $S$: Tracks both LIR (hot) blocks and resident HIR (cold) blocks
\item Queue $Q$: A FIFO queue of HIR blocks serving as eviction candidates
\end{itemize}

When an LIR block reaches the bottom of Stack $S$, "stack pruning" occurs to bound metadata overhead. LIRS consistently outperforms LRU-K and 2Q in trace-driven simulations, particularly for looping access patterns where the working set exceeds cache capacity—a scenario where LRU exhibits Bélády's anomaly behavior. MySQL 8.0's InnoDB storage engine uses LIRS-inspired policies \cite{jiang2002lirs}.

\subsection{ARC: Adaptive Self-Tuning}

Megiddo and Modha's ARC (Adaptive Replacement Cache) is celebrated for eliminating manual parameter tuning \cite{megiddo2003arc}. Unlike LRU-K (choosing $K$) or 2Q (sizing the probationary queue), ARC dynamically adjusts the balance between recency ($T_1$) and frequency ($T_2$) components.

ARC maintains four lists:
\begin{itemize}
\item $T_1$: Recent pages (accessed once recently)
\item $T_2$: Frequent pages (accessed multiple times)
\item $B_1$: Ghost entries for recently evicted $T_1$ pages
\item $B_2$: Ghost entries for recently evicted $T_2$ pages
\end{itemize}

The adaptation mechanism uses the ghost caches: a hit in $B_1$ suggests the recency component $T_1$ is undersized, increasing target parameter $p$. A hit in $B_2$ suggests the frequency component $T_2$ is undersized, decreasing $p$. This continuous online adaptation allows ARC to behave like LRU during workload transitions and like LFU during stable phases, achieving "empirical universality" across diverse workloads \cite{megiddo2003arc}.

Despite its effectiveness, IBM's patent on ARC forced open-source projects to seek alternatives. Bansal and Modha subsequently developed CAR (Clock with Adaptive Replacement), which achieves similar adaptivity without patent restrictions \cite{bansal2004car}. ARC is widely deployed in production systems including ZFS and various storage arrays.

Table I presents a comprehensive comparison of these classical algorithms across multiple dimensions.

\section{Production System Implementations}

Theory must withstand the harsh realities of concurrency, lock contention, and multi-core scalability. This section examines how major production systems implement buffer management.

\subsection{PostgreSQL: Clock Sweep with Partitioned Locking}

PostgreSQL eschews traditional LRU lists to avoid "LRU latch contention"—the scalability bottleneck arising from acquiring a central lock for every page access to maintain doubly-linked list ordering \cite{interdb2024, postgres_internals}.

Instead, PostgreSQL implements the Clock Sweep algorithm (also called Second Chance). Buffer descriptors form a circular array, with each descriptor containing a \texttt{usage\_count} field (integer, typically 0-5). When a buffer is accessed (pinned), its usage count is incremented atomically. During victim selection, a clock hand sweeps through the array; upon encountering a buffer with \texttt{usage\_count} $> 0$, the algorithm decrements the counter and continues. Only buffers with \texttt{usage\_count} $== 0$ and \texttt{pin\_count} $== 0$ (no active users) are evicted.

This approximates LRU while allowing $O(1)$ access operations without central locks. The mapping from (Tablespace, Database, RelFileNode, BlockNumber) to BufferID is protected by \texttt{BufMappingLock}, implemented as partitioned locks (default 128 partitions). This fine-grained locking enables concurrent I/O operations on different buffer pool regions \cite{postgres_internals}.

PostgreSQL also implements Ring Buffers for bulk operations (VACUUM, sequential scans, bulk inserts). These operations allocate small temporary rings (e.g., 256KB) within the shared buffer pool. Pages read into rings are immediately reused after processing, preventing maintenance tasks from polluting the main working set—a practical implementation of scan resistance at the access method layer \cite{rogov2021}.

\subsection{Oracle Database: Touch Count and Midpoint Insertion}

Oracle Database employs Touch Count LRU with Midpoint Insertion to mitigate list manipulation overhead \cite{oracle_buffer, lewis_oracle}. New blocks read from disk are inserted at the midpoint of the LRU list, not the head, effectively partitioning the list into "Hot" and "Cold" regions.

Each buffer header maintains a Touch Count (TCH) counter. Critically, accessing a buffer does not immediately move it to the head (which would require acquiring the \texttt{cache buffers lru chain} latch). Instead, the TCH increments. Promotion to the Hot region occurs only when: (1) TCH exceeds a configured threshold (typically 3-5), and (2) sufficient time has elapsed since the last increment, preventing single bursts from artificially inflating priority.

Buffer lookup is protected by Cache Buffers Chains (CBC) latches—fine-grained locks protecting hash bucket chains. High contention on specific CBC latches typically indicates "hot blocks"—individual data blocks concurrently accessed by thousands of sessions. Oracle's extensive wait event instrumentation allows DBAs to diagnose such contention patterns \cite{lewis_oracle}.

\subsection{Linux Kernel: Multi-Generational LRU and Refault Distance}

The Linux kernel memory management (MM) subsystem must handle diverse workloads without application hints. Historically, Linux used a simple two-list LRU (Active and Inactive lists). Recent kernels (5.18+) implement Multi-Generational LRU (MGLRU), which divides pages into multiple generations rather than binary active/inactive classification \cite{linux_mm}.

A key innovation is Refault Distance, introduced by Rik van Riel to address thrashing when working sets slightly exceed available memory \cite{linux_refault}. When evicting pages from the page cache, the kernel retains "shadow entries" in the radix tree (now xarray), storing eviction timestamps or sequence numbers.

Upon page fault, if a shadow entry exists, the kernel calculates the refault distance—the number of pages evicted between this page's eviction and its return. If refault distance is less than the Active list size, the page would have remained resident with a marginally larger cache. The kernel promotes such pages to the Active list and adjusts list sizes accordingly, implementing workingset-aware protection similar to LIRS but integrated into the OS virtual memory manager \cite{linux_refault}.

Table II compares implementation characteristics across these production systems.

\begin{table}[t]
\caption{Production System Buffer Management Comparison}
\label{tab:production}
\centering
\begin{tabular}{p{1.8cm}p{2.2cm}p{2cm}p{1.5cm}}
\toprule
\textbf{System} & \textbf{Algorithm} & \textbf{Concurrency} & \textbf{Scan Defense} \\
\midrule
PostgreSQL & Clock Sweep & Partitioned locks & Ring buffers \\
\midrule
Oracle & Touch Count LRU & CBC latches & Midpoint insertion \\
\midrule
MySQL InnoDB & Modified LRU & Mutex + RW locks & Young/Old lists \\
\midrule
Linux Kernel & MGLRU & Per-zone locks & Refault distance \\
\midrule
SQL Server & Clock & Spin locks & Lazy writer \\
\bottomrule
\end{tabular}
\end{table}

\section{Storage Media Evolution and Cost-Aware Buffering}

As storage technology evolved from HDDs (seek-dominated) to SSDs (wear-limited) and NVM (byte-addressable), buffer management algorithms adapted to accommodate physical asymmetries.

\subsection{Flash-Aware Policies}

Solid State Drives (SSDs) introduced two critical concerns: write amplification and finite program/erase cycles. Park et al.'s CFLRU (Clean-First LRU) adapts the replacement policy to prioritize clean page eviction over dirty pages \cite{park2006cflru}.

The rationale is straightforward: evicting a clean page incurs only a future read cost if the page is subsequently accessed. Evicting a dirty page requires an immediate write operation (flash program) plus a potential future read cost. CFLRU scans a window at the LRU end; if it finds clean pages, it evicts them preferentially. Only when the window contains exclusively dirty pages does the algorithm fall back to evicting the oldest dirty page.

This approach reduces write operations to the SSD, mitigating write amplification and extending device lifespan. Similar principles appear in FAB (Flash-Aware Buffer) and other flash-optimized policies \cite{park2006cflru}.

\subsection{NVM and Three-Tier Buffer Management}

Non-Volatile Memory (Intel Optane, 3D XPoint) introduced a tier between DRAM and SSD, offering persistence with near-DRAM latency (200-500ns) but lower bandwidth and higher cost than DRAM. Zhou et al.'s Spitfire addresses the three-tier DRAM-NVM-SSD hierarchy \cite{zhou2021spitfire}.

Spitfire recognizes that NVM's write bandwidth is significantly lower than DRAM. It employs a probabilistic migration policy: when reading a page from SSD, should it go to NVM or DRAM? Spitfire uses machine learning (simulated annealing) to dynamically tune these probabilities based on current workload characteristics (read/write ratio, access frequency distribution). This ensures NVM is not saturated by write-heavy pages that would be better served in DRAM or bypassed entirely \cite{zhou2021spitfire}.

\subsection{Pointer Swizzling: LeanStore}

In the era of NVMe arrays capable of millions of IOPS, the buffer manager itself becomes the bottleneck. Leis et al.'s LeanStore introduces pointer swizzling to buffer management \cite{leis2018leanstore}.

Traditional buffer managers use a hash table mapping (page ID) $\rightarrow$ (memory address). Every page access incurs hash lookup overhead, CPU cache misses, and latch acquisition. LeanStore "swizzles" pointers in parent pages to point directly to child page memory addresses. If a child page is evicted, the pointer is "unswizzled" back to a page ID.

This optimization removes the centralized buffer mapping table from the critical path for hot pages, achieving near-memory access speeds for in-buffer operations. LeanStore demonstrates that with modern storage performance, software overhead dominates, necessitating algorithmic innovations beyond replacement policy improvements \cite{leis2018leanstore}.

\section{Machine Learning and Learned Policies}

The static heuristics of classical algorithms are increasingly challenged by machine learning approaches that learn from access patterns and adapt to complex, non-linear workloads.

\subsection{Hardware-Level Learning for CPU Caches}

In CPU caches (L2/L3), replacement decisions occur in nanoseconds, requiring extremely lightweight predictors. Jain and Lin's Hawkeye formulates cache replacement as supervised learning \cite{jain2016hawkeye}.

Hawkeye reconstructs Bélády's optimal (MIN) algorithm from historical traces—MIN is provably optimal but requires knowledge of future accesses. Hawkeye trains a predictor to classify cache lines as "cache-friendly" (should be retained) or "cache-averse" (dead on arrival). Features include the Program Counter (PC) triggering the memory access, address patterns, and temporal characteristics.

The trained model maps PC values to cache-friendliness scores. At runtime, the hardware predictor makes eviction decisions based on these scores. Hawkeye demonstrates that learning from optimal solutions and distilling that knowledge into lightweight predictors can outperform traditional policies \cite{jain2016hawkeye}.

Shi et al.'s Glider applies deep learning (LSTMs) to cache replacement \cite{shi2019glider}. While LSTMs are too computationally intensive for real-time hardware decisions, Glider uses them offline to identify optimal features, then distills the model into a lightweight perceptron suitable for hardware implementation. This two-stage approach bridges the gap between ML model expressiveness and hardware constraints.

\subsection{Software-Level Reinforcement Learning}

At the OS and DBMS level, microsecond decision latencies permit more sophisticated machine learning. Vietri et al.'s LeCaR (Learning Cache Replacement) employs online learning with regret minimization \cite{vietri2018lecar}.

LeCaR maintains multiple "expert" policies (e.g., LRU and LFU). At each eviction decision, it selects a policy based on learned weights. After observing outcomes (cache hits/misses), LeCaR updates weights using regret—the cost of not following the best expert. This meta-learning approach dynamically adapts to workload characteristics, outperforming static ARC in small-cache scenarios where adaptivity is crucial \cite{vietri2018lecar}.

Rodriguez et al. extended LeCaR to CACHEUS, incorporating additional experts for scan resistance and churn resistance \cite{rodriguez2021cacheus}. CACHEUS handles four primitive workload patterns: recency-based, frequency-based, scan-heavy, and churn-heavy (rapid working set changes). The system learns optimal expert weights for each workload phase.

Beckmann et al.'s LHD (Least Hit Density) represents a probabilistic approach \cite{beckmann2018lhd}. Rather than heuristics, LHD calculates the conditional probability that an object will be hit within a time window, given its age and current request rate. It evicts objects with the lowest "hit density" (probability per byte). This fundamental shift from heuristic-based to probability-based management opens new avenues for incorporating statistical models and probabilistic reasoning.

Table III summarizes machine learning-based buffer management approaches.

\begin{table}[t]
\caption{Machine Learning-Based Buffer Management Approaches}
\label{tab:learned}
\centering
\begin{tabular}{p{1.5cm}p{2cm}p{2cm}p{1.5cm}}
\toprule
\textbf{Policy} & \textbf{ML Method} & \textbf{Features} & \textbf{Latency} \\
\midrule
Hawkeye \cite{jain2016hawkeye} & Supervised & PC, address & $<10ns$ \\
\midrule
Glider \cite{shi2019glider} & Deep learning & LSTM features & $ <10ns $ \\
\midrule
LeCaR \cite{vietri2018lecar} & Online learning & Expert regret & $<1\mu s$ \\
\midrule
CACHEUS \cite{rodriguez2021cacheus} & Multi-expert & 4 expert mix & $<1\mu s$ \\
\midrule
LHD \cite{beckmann2018lhd} & Probabilistic & Age, rate & $<1\mu s$ \\
\bottomrule
\end{tabular}
\end{table}

\section{Kernel Extensibility with eBPF}

A significant emerging trend in 2024-2025 is the use of eBPF (extended Berkeley Packet Filter) to enable application-specific memory management without kernel modifications.

\subsection{eBPF for Memory Management}

Traditionally, databases and operating systems operate in opposition—as Stonebraker described, databases bypass OS caching to avoid suboptimal decisions \cite{stonebraker1981}. eBPF offers a potential resolution by allowing userspace applications to inject custom logic into kernel memory management paths \cite{mores2024ebpf}.

Mores et al.'s eBPF-mm framework attaches BPF programs to the kernel's page reclaim path (\texttt{shrink\_page\_list}) \cite{mores2024ebpf}. When the kernel needs to reclaim memory under pressure, it invokes the attached BPF program. The application-provided program can inspect page metadata (inode, mapping, access patterns) and return verdicts: PROTECT (keep in memory), EVICT (free this page), or PASS (use default kernel policy).

This mechanism enables cross-layer semantic passing: the DBMS can inform the kernel about high-value pages (index roots, catalog pages) versus low-value pages (temporary data, scan results). The kernel makes informed decisions without the DBMS completely bypassing OS caching, potentially achieving better global memory utilization \cite{mores2024ebpf}.

\subsection{Application-Specific Virtual Memory}

Jalalian et al.'s ExtMEM extends eBPF principles to virtual memory management \cite{jalalian2024extmem}. ExtMEM allows data-intensive applications (graph processing, scientific computing) to manage their own swap policies and target devices. Applications can implement custom eviction policies via BPF programs and direct swapped pages to application-specific backing stores rather than the generic Linux swap partition.

This is particularly valuable for workloads with known access patterns that differ from the kernel's assumptions. For example, graph algorithms with predictable traversal patterns can implement prefetching and eviction strategies tailored to their specific needs, bypassing the generic LRU-based swap system that often causes thrashing \cite{jalalian2024extmem}.

\section{Disaggregated Memory and Future Architectures}

The most recent architectural shift is toward memory disaggregation, where compute nodes access a pool of memory nodes via high-speed interconnects like RDMA.

\subsection{Fine-Grained Disaggregated Memory}

Wang et al.'s FineMem addresses a fundamental challenge in disaggregated memory: granularity mismatch \cite{wang2025finemem}. RDMA operations are most efficient with large transfers (64KB-1MB), but databases require fine-grained access (4KB pages). Naively using RDMA for 4KB operations wastes bandwidth and adds latency.

FineMem introduces a lock-free allocator managing remote memory at fine granularity without RPC overhead. It uses one-sided RDMA reads/writes with careful cache coherence protocols, achieving near-local-memory performance for remote accesses \cite{wang2025finemem}.

\subsection{Scalability for Many-Core Systems}

Liu et al.'s ScaleCache targets the scalability challenges of modern many-core systems (64+ cores) \cite{liu2025scalecache}. Traditional buffer managers use centralized latches protecting the buffer mapping table and LRU lists, creating severe contention bottlenecks.

ScaleCache employs per-core buffer pools with lock-free coordination protocols. Each core manages its own buffer region, minimizing cross-core communication. A hierarchical buffer organization (per-core $\rightarrow$ per-NUMA-node $\rightarrow$ global) balances local optimization with system-wide efficiency \cite{liu2025scalecache}.

\subsection{CXL and Tiered Memory Management}

Compute Express Link (CXL) is an emerging interconnect standard enabling memory devices to be attached to CPUs with cache-coherent, low-latency access. Zhong et al.'s Memstrata explores buffer management in CXL-enabled systems \cite{zhong2024memstrata}.

Memstrata manages a three-tier hierarchy: local DRAM (50ns), CXL-attached memory (200ns), and local SSD $(10 \mu s)$. The system monitors access patterns and migrates hot pages to faster tiers while demoting cold pages to slower tiers. The key challenge is balancing migration overhead against access latency improvements, requiring sophisticated cost modeling \cite{zhong2024memstrata}.

\section{Research Directions}

Based on our survey, we identify several promising research directions for next-generation buffer management systems.

\subsection{Challenges and Open Problems}

Despite decades of research, three fundamental challenges persist:

\textbf{C1: Cross-Layer Opacity.} DBMS and OS manage memory independently, unable to share semantic information efficiently. Current bypass mechanisms (O\_DIRECT) sacrifice potential OS optimizations like readahead and system-wide pressure coordination.

\textbf{C2: Static Policy Selection.} Production systems typically commit to a single replacement policy at design time. However, workloads exhibit temporal heterogeneity—OLTP transactions, analytical scans, and maintenance tasks benefit from different policies.

\textbf{C3: Heterogeneous Memory Complexity.} Classical algorithms assume uniform access costs. Modern systems with DRAM-CXL-NVM-SSD tiers require tier-aware placement and migration strategies that account for diverse latency, bandwidth, and cost characteristics.

\subsection{Proposed Research Direction}

We propose investigating adaptive buffer management frameworks that integrate machine learning with eBPF-based kernel extensibility. Figure 1 illustrates a conceptual architecture with four layers:

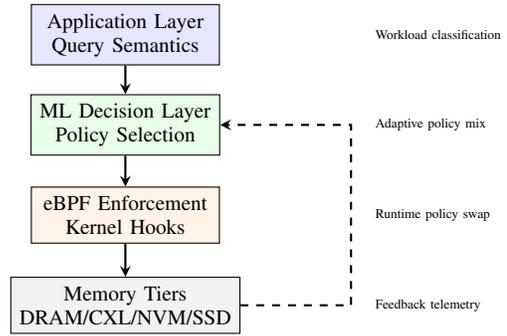
\begin{figure}[t]
\centering
\begin{tikzpicture}[
    node distance=0.6cm,
    box/.style={rectangle, draw, minimum width=2.5cm, minimum height=0.7cm, align=center, font=\footnotesize},
    arrow/.style={->, >=stealth, thick}
]

\node[box, fill=blue!10] (app) at (0,0) {Application Layer\\Query Semantics};
\node[box, fill=green!10] (ml) at (0,-1.2) {ML Decision Layer\\Policy Selection};
\node[box, fill=orange!10] (ebpf) at (0,-2.4) {eBPF Enforcement\\Kernel Hooks};
\node[box, fill=gray!10] (mem) at (0,-3.6) {Memory Tiers\\DRAM/CXL/NVM/SSD};

\draw[arrow] (app) -- (ml);
\draw[arrow] (ml) -- (ebpf);
\draw[arrow] (ebpf) -- (mem);
\draw[arrow, dashed] (mem) -- (3,-3.6) -- (3,-1.2) -- (ml);

\node[anchor=west, font=\tiny, align=left] at (3.2,0) {Workload classification};
\node[anchor=west, font=\tiny, align=left] at (3.2,-1.2) {Adaptive policy mix};
\node[anchor=west, font=\tiny, align=left] at (3.2,-2.4) {Runtime policy swap};
\node[anchor=west, font=\tiny, align=left] at (3.2,-3.6) {Feedback telemetry};

\end{tikzpicture}
\caption{Conceptual architecture for adaptive buffer management integrating ML-based policy selection with eBPF enforcement across heterogeneous memory tiers.}
\label{fig:architecture}
\end{figure}

\textbf{Layer 1: Application Semantic Layer.} The DBMS provides workload context (query types, access patterns, scan detection) to enable informed policy decisions.

\textbf{Layer 2: ML Decision Layer.} Machine learning models predict page reuse patterns and select optimal replacement policies dynamically. Multi-armed bandit approaches (e.g., Thompson Sampling) can balance exploration of new policies against exploitation of known-good policies.

\textbf{Layer 3: eBPF Enforcement Layer.} Kernel hooks at memory management points (page reclaim, page fault) execute BPF programs implementing selected policies. This enables policy changes without kernel recompilation.

\textbf{Layer 4: Heterogeneous Memory Tiers.} Unified management of DRAM, CXL, NVM, and SSD with tier-aware placement based on access frequency predictions and tier characteristics.

Key research questions include: (1) How to efficiently extract and communicate semantic information from DBMS to kernel? (2) What ML architectures can predict reuse patterns with $<1\mu s$ latency? (3) How to coordinate tier migration decisions across multiple concurrent workloads? (4) What are the appropriate abstractions for application-kernel cooperation?

\subsection{Future Directions}

Beyond the adaptive framework, several specific directions warrant investigation:

\textbf{D1: Query-Plan-Aware Buffer Management.} Integrate query optimizer statistics and execution plans into buffer management decisions. For example, knowing a nested-loop join will repeatedly scan the inner table enables proactive retention policies.

\textbf{D2: Energy-Efficient Tiering.} Extend cost models to include power consumption. CXL memory may use less power than DRAM, enabling performance-per-watt optimizations in energy-constrained environments.

\textbf{D3: Multi-Tenant Fairness.} Develop fairness-aware policies for cloud databases ensuring no single tenant monopolizes fast memory tiers. This requires mechanisms for resource quotas and priority-based allocation.

\textbf{D4: Hardware-Software Co-Design.} Explore how buffer management principles can inform hardware cache controller design, potentially through custom ISA extensions enabling semantic hints from software to hardware caches.

\textbf{D5: Federated Learning for Policy Transfer.} Train ML models across diverse deployments, enabling transfer learning for new installations without extensive local trace collection.

\section{Conclusion}

This survey has traced the evolution of buffer management across four decades, from Stonebraker's critique of OS-DBMS integration through the development of sophisticated algorithms (LRU-K, LIRS, ARC) to contemporary machine learning and disaggregated memory architectures.

Several clear trends emerge from our analysis:

\textbf{From Static to Adaptive.} Early algorithms used fixed heuristics (LRU, LFU). Modern approaches like ARC introduced self-tuning, and recent ML-based policies (LeCaR, CACHEUS) enable continuous adaptation to workload changes.

\textbf{From Homogeneous to Heterogeneous.} Classical algorithms assumed uniform storage. Contemporary systems must manage complex memory hierarchies (DRAM-CXL-NVM-SSD) with diverse performance characteristics.

\textbf{From Isolation to Cooperation.} The historical OS-DBMS divide is being bridged by technologies like eBPF, enabling cross-layer optimization without sacrificing modularity.

\textbf{From Heuristic to Learned.} Machine learning is transitioning buffer management from hand-crafted policies to data-driven approaches that learn from access patterns.

Looking forward, successful buffer management systems will likely combine multiple innovations: learned policies for workload adaptation, eBPF for cross-layer cooperation, and tier-aware placement for heterogeneous memory. The challenges of managing increasingly complex memory hierarchies in cloud-native, multi-tenant environments present rich opportunities for continued research.

As memory technologies continue to evolve (CXL, computational storage, processing-in-memory) and workloads grow more diverse (OLTP, OLAP, AI inference, graph analytics), intelligent buffer management remains as critical today as when Stonebraker identified the problem four decades ago. The field continues to offer fertile ground for systems research combining theory, implementation, and machine learning.

\end{document}